\documentclass[9pt,twocolumn,twoside]{optica}
\usepackage[labelsep=colon]{caption}
\setboolean{shortarticle}{false}

\title{Observation of resonance fluorescence and the Mollow-triplet from a coherently driven site-controlled quantum dot}

\author[1,+]{Sebastian Unsleber}
\author[1,+]{Sebastian Maier}
\author[2]{Dara P. S. McCutcheon}
\author[1,3]{Yu-Ming He}
\author[1]{Michael Dambach}
\author[4]{Manuel Gschrey}
\author[2]{Niels Gregersen}
\author[2]{Jesper M\o rk}
\author[4]{Stephan Reitzenstein}
\author[1,3,5]{Sven H\"ofling}
\author[1,*]{Christian Schneider}
\author[1]{Martin Kamp}

\affil[1]{Technische Physik and Wilhelm Conrad R\"ontgen Research Center for Complex Material Systems, Physikalisches Institut,
Universit\"at W\"urzburg, Am Hubland, D-97074 W\"urzburg, Germany}
\affil[2]{Department of Photonics Engineering, Technical University of Denmark, \O rsteds Plads, 2800 Kgs. Lyngby, Denmark}
\affil[3]{Hefei National Laboratory for Physical Sciences at the Microscale and Department of Modern Physics,
$\&$ CAS Center for Excellence and Synergetic Innovation Center in Quantum Information and Quantum Physics,
University of Science and Technology of China, Hefei, Anhui 230026, China}
\affil[4]{Institut f\"ur Festk\"orperphysik, Technische Universit\"at Berlin, Hardenbergstraße 36, 10623 Berlin, Germany}
\affil[5]{SUPA, School of Physics and Astronomy, University of St Andrews, St Andrews, KY16 9SS, United Kingdom}

\affil[+]{These authors contributed equally}
\affil[*]{Corresponding author: christian.schneider@physik.uni-wuerzburg.de}

\dates{Compiled \today}

\ociscodes{(230.5590) Quantum-well, -wire and -dot devices; (220.4241) Nanostructure fabrication;  (270.0270) Quantum optics; (300.6470) Spectroscopy, semiconductors}

\doi{\url{http://dx.doi.org/10.1364/optica.XX.XXXXXX}}

\begin{abstract}

Resonant excitation of solid state quantum emitters has the potential to deterministically excite a localized exciton while ensuring a maximally coherent emission. In this work, we demonstrate the coherent coupling of an exciton localized in a lithographically positioned, site-controlled semiconductor quantum dot to an external resonant laser field. For strong continuous-wave driving we observe the characteristic
Mollow triplet and analyze the Rabi splitting and sideband widths as a function of driving strength
and temperature. The sideband widths increase linearly with temperature and the square of the driving strength, which we explain via coupling of the exciton to longitudinal acoustic phonons. We also find an increase of the Rabi splitting with temperature, which indicates a temperature induced delocalization of the excitonic wave function resulting in an increase of the oscillator strength. Finally, we demonstrate coherent control of the exciton excited state population via pulsed resonant excitation and observe a damping of the Rabi oscillations with increasing pulse area, which is consistent with our exciton–photon coupling model. We believe that our work outlines the possibility to implement fully scalable platforms of solid state quantum emitters. The latter is one of the key prerequisites for more advanced, integrated nanophotonic quantum circuits.

\end{abstract}

\begin{document}

\maketitle
\thispagestyle{fancy}
\ifthenelse{\boolean{shortarticle}}{\abscontent}{}

\section{Introduction}

Semiconductor self-assembled quantum dots (QDs) are prime candidates for solid state quantum emitters~\cite{Michler2000,Santori02}. Many groundbreaking experiments have shown this great potential, e.g. in quantum key distribution experiments~\cite{Waks2002,Heindel2012}, the demonstration of spin--photon entanglement~\cite{DeGreve2012,Gao2012}, and the emission of highly indistinguishable single photons~\cite{He2013,Wei2014a}. Furthermore, quantum dots offer advantages compared to alternative platforms, e.g. cold atoms or trapped ions, such as their ability to be electrically contacted~\cite{Heindel2010,Ellis2008,Yuan2002}, and the possibility to be implemented into photonic architectures or networks~\cite{Hoang12,Yao09,Joens2015}.

Thus far,  most relevant demonstrations were carried out based on randomly positioned QDs. Scalable schemes for the implementation of solid state quantum bits or quantum emitters, however, require position control over the QDs. This has triggered extensive research activities to realize site-controlled QD (SCQD) arrays~\cite{Ishikawa2000,Schmidt2007,Schneider2009}. While the demonstration of single photon emission~\cite{Baier2004, Schneider2009a}, emission of indistinguishable \cite{Joens2013} and polarization entangled photons~\cite{Juska2013} in this system has been demonstrated, the implementation of resonant fluorescence in this system remains elusive. Resonant coupling of a laser to the quantum emitter, however, is key towards the emission of single photons with high indistinguishability and to coherently control the state of the excitonic qubit. In this work, we demonstrate for the first time the resonant excitation of a SCQD. We observe the characteristic Mollow triplet~\cite{Ates2009a,Flagg2009, Vamivakas2009} in the resonance fluorescence spectra under continuous wave excitation conditions, and demonstrate the coherent control of the excited state SCQD population via pulsed resonant excitation. Furthermore, we analyze the light matter coupling for varying temperatures and driving strengths, which allows us to characterize the strength of exciton--phonon interactions and assess the lateral exciton extension in the quantum dot.

\section{Sample structure and setup}
\label{sec:Sample structure and setup}

\begin{figure}[t!]
\begin{center}
\includegraphics[width=0.48\textwidth]{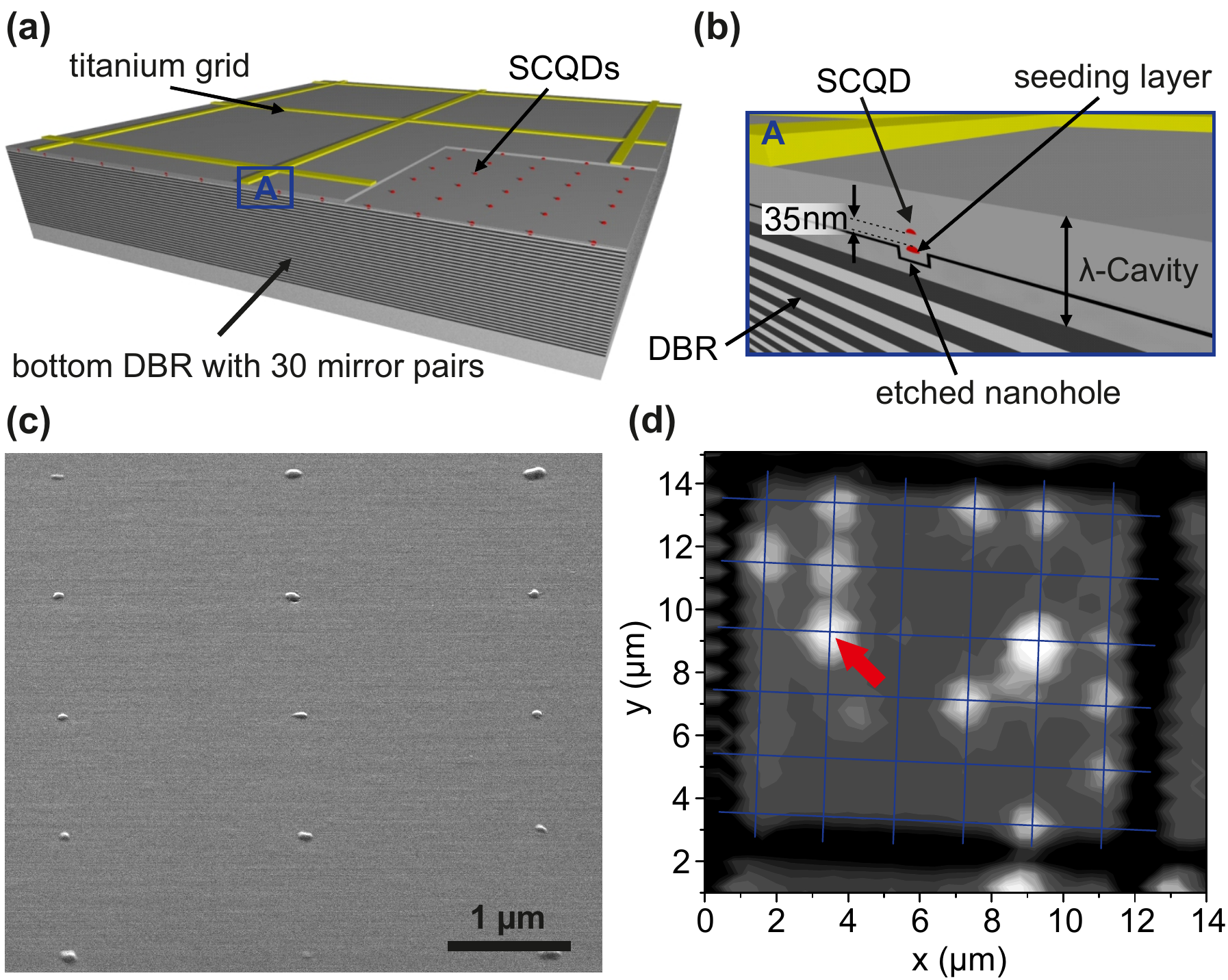}
\caption{(a) Schematic drawing of the sample structure showing the lower DBR and the $\lambda$-thick spacer with embedded SCQDs. 
 (b)~Detailed layer sequence with two quantum dot layers and the $35$~nm thick GaAs separation 
layer in between. (c)~SEM image of SCQDs on an uncapped calibration sample. (d)~$\mu$PL-map of a $14\times 14~\mu\text{m}^2$ area. 
The bright spots are SCQDs which are located on a 2~$\mu$m grid which was predefined by nanoholes. The blue grid serves as a guide to the eye, the red arrow indicates the SCQD which is subject to the in depth study.}
\label{Fig1}
\end{center}
\end{figure}

Our sample consists of stacked site-controlled InAs quantum dots which are embedded in a single-sided planar cavity to improve the light extraction, as shown in Fig.~\ref{Fig1}(a). After the growth of the bottom distributed Bragg reflector (DBR) with $30$ quarter-wavelength thick AlAs/GaAs mirror pairs and a $85$ nm thick GaAs layer via molecular beam epitaxy, arrays of nanoholes with a $2~\mu$m pitch are defined on the wafer by means of electron beam lithography and wet-chemical etching~\cite{Schneider2009}. The structure design in principle allows to yield photon extaction efficiencis on the order of $\approx10-12~\%$~\cite{Royo2001} for the microscope objective with $NA=0.42$ we used. 
We note that this number can be further increased by utilising micropillar cavities (based on devices with top DBR~\cite{Gazzano2013,Heindel2010,Schneider2009a}, or by shaping the confinement in broadband approaches via dielectric lenses \cite{Maier2014,Gschrey2015,Sapienza2015}. The pre-patterned sample is deoxidized with thermally activated hydrogen before the second epitaxial growth step is performed. 
The nanoholes are then capped by $10$ nm GaAs and a first InAs quantum dot layer (seeding layer) at a substrate temperature of $540~^{\circ}$C to maximize the migration length and to ensure that nucleation preferentially takes place in the nanoholes. 
To improve the optical quality of the quantum dots, a second layer of InAs quantum dots is separated by a $35$ nm thick GaAs separation layer, while the positioning of the SCQDs is maintained by the vertical strain field (Fig.~\ref{Fig1}(b)). 
In Fig.~\ref{Fig1}(c) a scanning electron microscopy (SEM) image shows SCQDs with a $2~\mu$m period on an uncapped reference sample, recorded under an angle of $70^{\circ}$ to enhance the imaging contrast.  The emission wavelength range of the buried QDs in our device is shifted towards $900$~nm by performing an in-situ annealing step, before a $131$ nm thick GaAs layer completes the structure. 
A titanium grid on the surface of the sample with a pitch of $12~\mu$m and $300$ nm width (Fig.~\ref{Fig1}(a)) is defined by means of e-beam lithography. This serves as a coordinate system for orientation, with each grid square containing a regular array of 36 SCQDs.

The sample is mounted on the cold-finger of a He flow cryostat and is excited by a continuous wave laser at 532~nm. Fig.~\ref{Fig1}(d) shows a $14\times 14~\mu$m$^2$ large array scan of the photoluminescence (PL)-signal of the sample. The bright spots between the titanium frames can be attributed to the PL signal of SCQDs which are located on the expected positions of the array. SEM images of the uncapped sample shown in Fig.~\ref{Fig1}(c), indicate that SCQDs are present at all expected positions. As such, we attribute dark spots on the grid to a large variation in quantum efficiencies  ~\cite{Albert2010}. The SCQD indicated with a red arrow in Fig.~\ref{Fig1}(d) is the basis of detailed subsequent  $\mu$PL investigations. The QD is excited either by a fiber coupled continuous wave diode laser or a pulsed Ti:Sapphire laser with a repetition rate of $f=82~$MHz and a pulse length of $\Delta t\approx1.2~$ps. In order to suppress the resonant laser, we use a cross polarization configuration where the excitation laser and the detected SCQD signal have perpendicular polarizations. The emitted photons are then analyzed via a double monochromator with two 
$1200~\mathrm{lines}/\mathrm{mm}$ gratings and a nitrogen cooled Si-CCD ($\Delta E_{\mathrm{Res}}\approx 20~\mu$eV).

\section{Theory}
\label{sec:Theory}

For the interpretation of our experimental results, we employ basic concepts of the theory of a resonantly driven quantum dot coupled to longitudinal acoustic (LA) 
phonons. In Ref.~\cite{McCutcheon2013} it was shown that provided $\Omega<k_{B} T<\omega_{c}$, with $\Omega$ the Rabi frequency, $k_B$ the Boltzmann constant, $T$ the sample temperature, and $\omega_{c}$ the cut-off frequency of the phonon environment, the effect of phonons on the excitonic dynamics can 
be accurately captured by a renormalization of the bare Rabi frequency and the introduction of a phonon-induced pure-dephasing rate. We write the optical Bloch equations describing the excitonic degrees of freedom as 
\begin{align}
\dot{\alpha}_x&=-\Gamma_2\,\alpha_x,\label{alphaxdot}\\
\dot{\alpha}_y&=-\Gamma_2\,\alpha_y - \Omega_r \alpha_z,\label{alphaydot}\\
\dot{\alpha}_z&=-\Gamma_1\,\alpha_z + \Omega_r \alpha_z-\Gamma_1,\label{alphazdot}
\end{align}
where $\alpha_i=\left\langle\sigma_i\right\rangle=\mathrm{Tr}(\rho \sigma_{i})$ for $i=x,y,z$ and $\rho$ the exciton density operator, 
and we work in a basis where $\sigma_{x}=\left|e\right\rangle\left\langle g\right|+\left|g\right\rangle\left\langle e\right|$, $\sigma_{y}=-i\left|e\right\rangle\left\langle g\right|+i\left|g\right\rangle\left\langle e\right|$, $\sigma_{z}=\left|e\right\rangle\left\langle e\right|-\left|g\right\rangle\left\langle g\right|$ with 
$\left|e\right\rangle$ and $\left|g\right\rangle$ the single exciton and ground states respectively. The total dephasing rate is given by 
$\Gamma_{2}=\frac{1}{2}\Gamma_{1}+\gamma_{PD}+\gamma_{0}$ with 
$\Gamma_{1}=1/T_{1}$ spontaneous emission rate, $\gamma_{0}$ captures dephasing not attributed to phonons, 
and $\gamma_{PD}$ is the phonon induced dephasing rate, which we define below. The effective Rabi frequency 
is given by (for $\hbar=1$)
\begin{equation}
\Omega_{r}=\mu E_{0} R~(=\kappa\sqrt{P}),
\label{Omegar}
\end{equation}
with $\mu$ the dipole moment of the emitter, $E_{0}$ the electric field strength of the laser light, $R$ a phonon induced renormalization factor, $\kappa$ representing the product $\mu R$ and $P$ the laser power. 

Exciton--phonon coupling is characterized by the spectral density, which for coupling to LA phonons takes the form 
$J(\omega)=\alpha \omega^{3} \exp[-(\omega/\omega_{c})^{2}]$, with $\alpha$ capturing the overall strength of the interaction~\cite{Ramsay2010a,Ramsay2010}. 
In terms of the spectral density, 
the phonon-induced renormalisation factor is given by~\cite{McCutcheon2010,McCutcheon2013,Wei2014}
\begin{equation}
R=\exp\bigg[-\frac{1}{2}\int_0^{\infty}\mathrm{d}\omega \frac{J(\omega)}{\omega^2}\coth(\omega/2k_B T)\bigg].
\label{Bav}
\end{equation}
In the weak exciton--phonon coupling limit, the phonon induced pure-dephasing rate 
is given by
\begin{equation}
\gamma_{PD}=(\pi/2)J(\Omega_r)\coth(\Omega_r/2k_B T),
\label{gammaPDFull}
\end{equation}
which becomes 
\begin{equation}
\gamma_{PD}=\pi \alpha k_B T \Omega_r^2~(=\chi \Omega_r^2),
\label{gammaPD}
\end{equation}
provided $\Omega_r\ll k_B T,\omega_c$ and $\chi$ being the product $\pi \alpha k_B T $. We note that this last condition is typically met in continuous wave measurements due to 
the relatively small Rabi frequencies. For pulsed measurements, however, instantaneous Rabi frequencies can reach levels comparable to $\omega_c,k_B T \sim 1~\mathrm{meV}$ (at $T=10~\mathrm{K}$). In the following, we therefore use the simpler expression given in Eq.~({\ref{gammaPD}}) for the continuous wave measurements, but the full expression in Eq.~({\ref{gammaPDFull}}) when investigating pulsed excitation conditions.

The incoherent component of the resonance fluorescence spectrum is given by $S(\omega)\propto \mathrm{Re}[\int_0^{\infty}\mathrm{d}\tau (g^{(1)}(\tau)-g^{(1)}(\infty))\exp[-i\omega\tau]]$, 
where $g^{(1)}(\tau)=\lim_{t\to\infty}\left\langle \sigma^{\dagger}(t+\tau)\sigma(t)\right\rangle$ with $\sigma=\left|g\right\rangle\left\langle e\right|$ is the steady-state first order field correlation 
function. From Eqs.~({\ref{alphaxdot}}) to ({\ref{alphazdot}}) calculation of the field correlation function and fluorescence spectrum proceeds by invoking the quantum regression theorem~\cite{carmichael,mccutcheon2015}. It is found that above saturation ($\Omega_{r}\gg \Gamma_1$), the fluorescence spectrum consists of three Lorentzian peaks, one centred at the laser driving frequency, and two more positioned on either side at a distance of $\Omega_{r}$. In this regime the full width at half maximum (FWHM) of the sidebands is given by~\cite{Wei2014}
\begin{equation}
\Delta \omega = \frac{3}{2}\Gamma_1+\gamma_{PD}+\gamma_0.
\label{SidebandWidths}
\end{equation}
From these expressions we see that we expect the sideband widths to increase linearly with $\Omega_{r}^{2}$, with an intersect (as $\Omega_{r} \to 0$)  given by $\smash{(3/2)\Gamma_1+\gamma_0}$. Moreover, since $\smash{E_0\sim\sqrt{P}}$, for constant $\mu$ one expects the gradient $(\mathrm{d}\Omega_r/\mathrm{d}\sqrt{P})\sim \mu R$ to decrease with temperature since $R$ decreases, as was experimentally observed in Ref.~\cite{Wei2014}.

\section{Experimental Results and Analysis}
\label{sec:Experimental Results and Analysis}

\begin{figure}
\begin{center}
\includegraphics[width=0.48\textwidth]{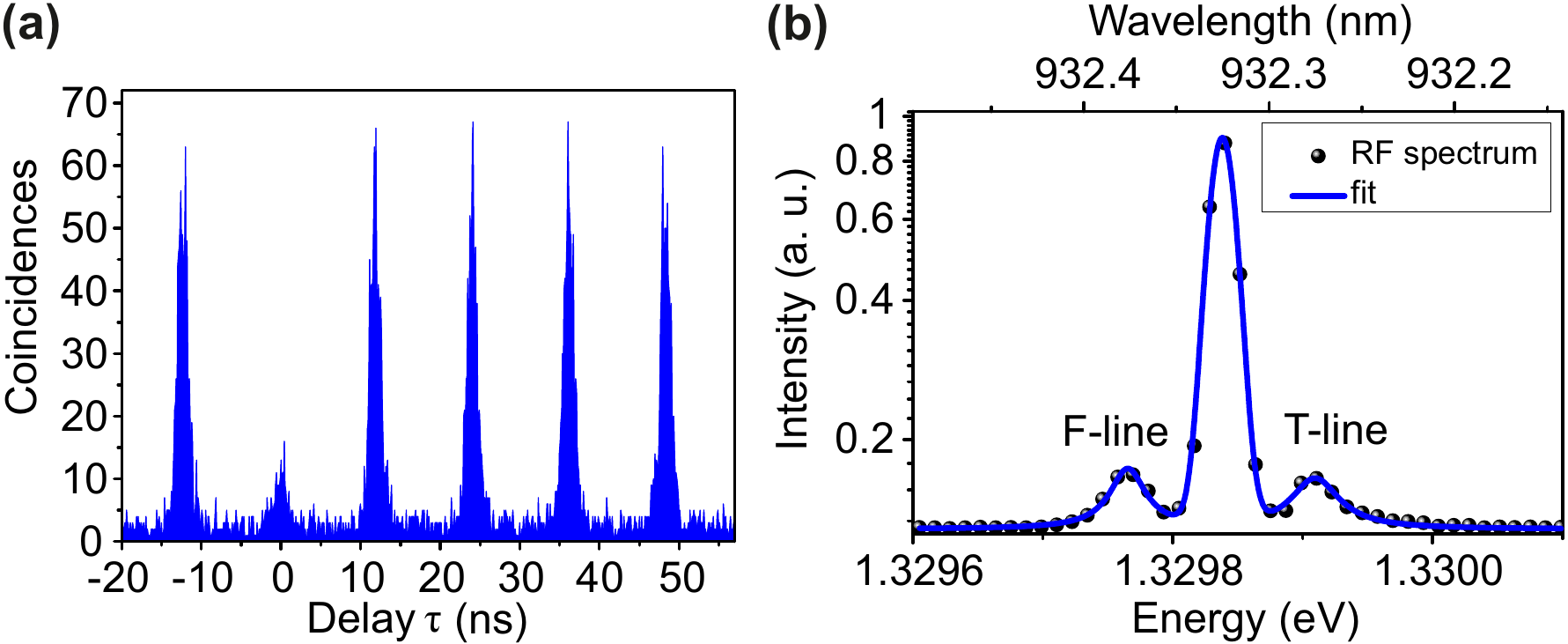}
\caption{(a) Second order autocorrelation function for pulsed non-resonant excitation and a sample temperature of $T=6.8$~K. We extract a $g^{(2)}(0)$-value of $g^{(2)}(0)=0.39\pm0.02$. (b)~Resonance fluorescence (RF) spectrum of a SCQD for a pump power of $P=746~$nW at $T=5~$K. The triplet peak structure is a signature of coherent coupling between the quantum dot exciton and the laser field.} 
\label{Fig2}
\end{center}
\end{figure}

First, we study the emission properties of the SCQD under pulsed nonresonant excitation. To do so, we couple the spectrally filtered microphotoluminescence emission into a fiber-based Hanbury Brown and Twiss setup in order to measure the second order autocorrelation function. A histogram of coincidence events measured as a function of detection time delay is shown in Fig.~\ref{Fig2}(a). The suppressed peak around $\tau\approx0$~ns is a clear signature for single photon emission from our SCQD. We extract a $g^{(2)}(0)$-value by dividing the area of the central peak by the average area of the surrounding peaks, which yields a value of $g^{(2)}(0)=0.39\pm0.02$ (corrected according to \cite{Michler2002} by spectral background emission $g^{(2)}(0)=0.22\pm0.02$). Remaining two-photon detection events arise from refilling of the SCQD due to the non-resonant excitation. In addtition, fitting the sidepeaks with a two-sided exponential function allows us to extract the lifetime of the exciton and we find $T_1=(561\pm68)$~ps. Compared to standard In(Ga)As QDs grown under comparable conditions~\cite{Johansen2008}, the $T_1$-time is slightly reduced. We believe this is a consequence of the presence of non-radiative recombination channels which play a significant role for QDs grown on patterned substrates~\cite{Albert2010}.

Next, we study the SCQD's emission properties under strict resonant excitation. Therefore, we drive the single exciton transition on resonance with a continuous wave laser, and analyze the spectrum of 
emitted radiation. Fig.~\ref{Fig2}(b) shows a representative spectrum of resonance fluorescence from the investigated SCQD corresponding to a pump power of $P=746~$nW and at a temperature of $T=5~$K. Due to imperfections of the sample surface the laser is not fully suppressed, resulting in a ratio of the sidepeak area to the central peak of approximately $1:12$. We fit the spectrum to a sum of two Lorenztians for the sidepeaks and a Gaussian for the central peak, which is resolution limited. From the fit, for this pump power and temperature we obtain sidepeak widths of $\Delta\omega_F=(31\pm7)~\mu$eV and $\Delta\omega_T=(43\pm9)~\mu$eV and a Rabi splitting of $\Omega_r=72~\mu$eV. 

\begin{figure}
\begin{center}
\includegraphics[width=0.48\textwidth]{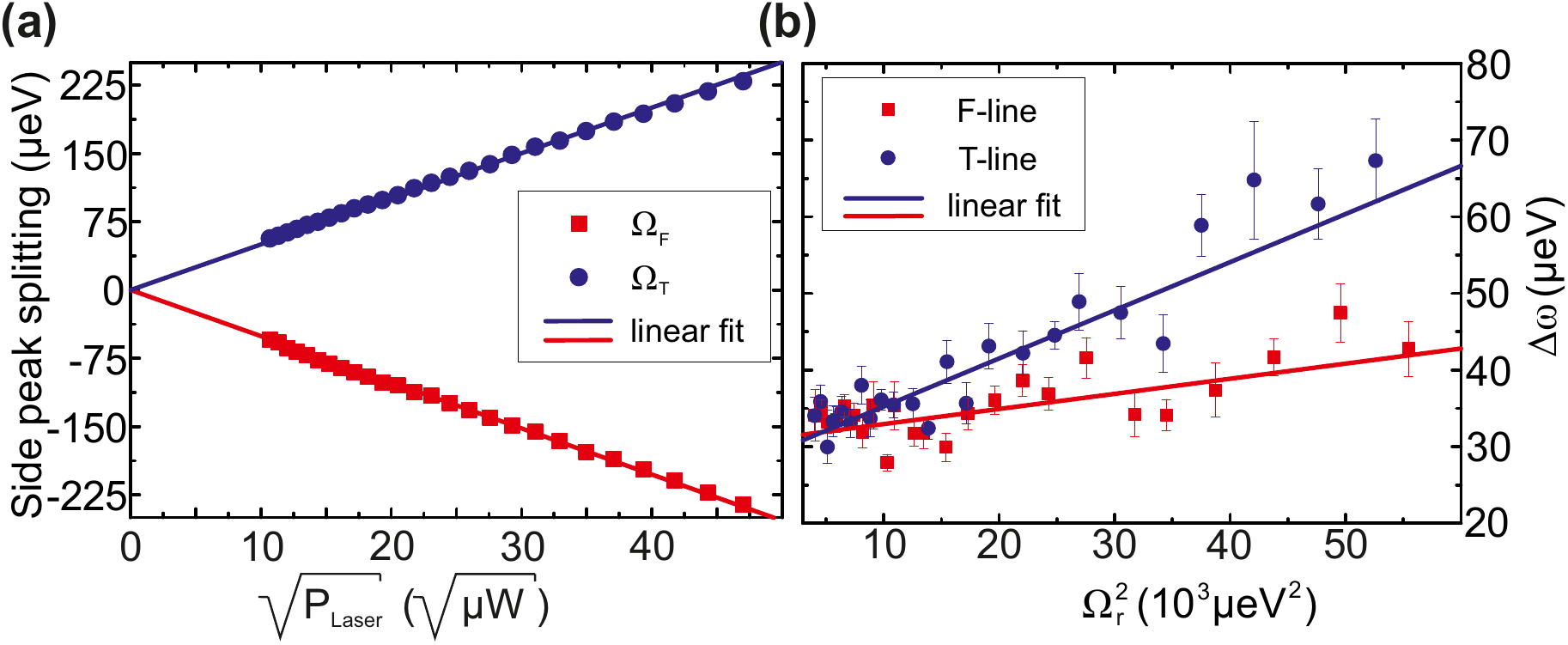}
\caption{(a)~Rabi splitting as a function of the square root of the pump power at $T=5~$K. A clear linear increase is observed with effective dipole moment of $\kappa=\pm(5.04\pm0.01)~\frac{\mu\text{eV}}{\sqrt{\mu\text{W}}}$. (b) Linewidth of the Mollow sidepeaks as a function of the Rabi frequency. A linear trend is consistent with Eq.~({\ref{gammaPD}}) suggesting dephasing caused by coupling to LA phonons.}
\label{Fig3}
\end{center}
\end{figure}

For increasing pump power but at fixed temperature, the sidepeak splitting increases as predicted by theory. Fig.~\ref{Fig3}(a) shows this splitting as a function of the square root of the laser power. 
The splitting follows a clear linear dependence, as is expected from Eq.~({\ref{Omegar}}) and recalling that the field amplitude $E_0\propto\sqrt{P}$~\cite{Vamivakas2009}. From the fits, we obtain a slope of $\kappa=\pm(5.04\pm0.01)~\mu\text{eV}/\sqrt{\mu\text{W}}$, which is slightly smaller than it is reported for self-organized QDs (\cite{Wei2014}:$\kappa=7.6~\mu\text{eV}/\sqrt{\mu\text{W}}$). 
Fig.~{\ref{Fig3}}(b) shows the variation of the sideband linewidths as a function of the square of the Rabi frequency. We observe a linear behaviour with gradients of $\chi_F=(2.0\pm0.6)\cdot10^{-4}~\mu\text{eV}^{-1}$ for the F-line and $\chi_T=(6.3\pm0.7)\cdot10^{-4}~\mu\text{eV}^{-1}$ for the T-line. The linear behaviour is consistent with Eq.~({\ref{gammaPD}}), suggesting excitation induced dephasing 
caused by coupling to LA phonons in our sample. Additionally, from Eq.~({\ref{SidebandWidths}}) we see the intersects with the y-axis give non-phonon induced dephasing rates of $\gamma_{0}\approx 30~\mu\mathrm{eV}$. We note that for resonant excitation, our theory predicts F- and T-sidebands of equal width. We attribute the systematic asymmetry to a slight detuning of the QD level from the excitation laser, which becomes more pronounced at the high pump powers we use in order to maximize the Rabi splitting. Such asymmetries have been predicted in \cite{Ulhaq2013} for systems with relatively high dephasing rates, and arise as a result of the exciton-phonon coupling interaction producing behaviour which departs from a simple pure-dephasing model. The larger broadening of the T-line seen in our measurements suggests our laser is blue shifted from the QD transition. A systematic investigation of these off-resonance effects and their underlying physical mechanisms has yet to be performed, and this is a topic which we plan to explore fully in future work.

\begin{figure}[b!]
\begin{center}
\includegraphics[width=0.48\textwidth]{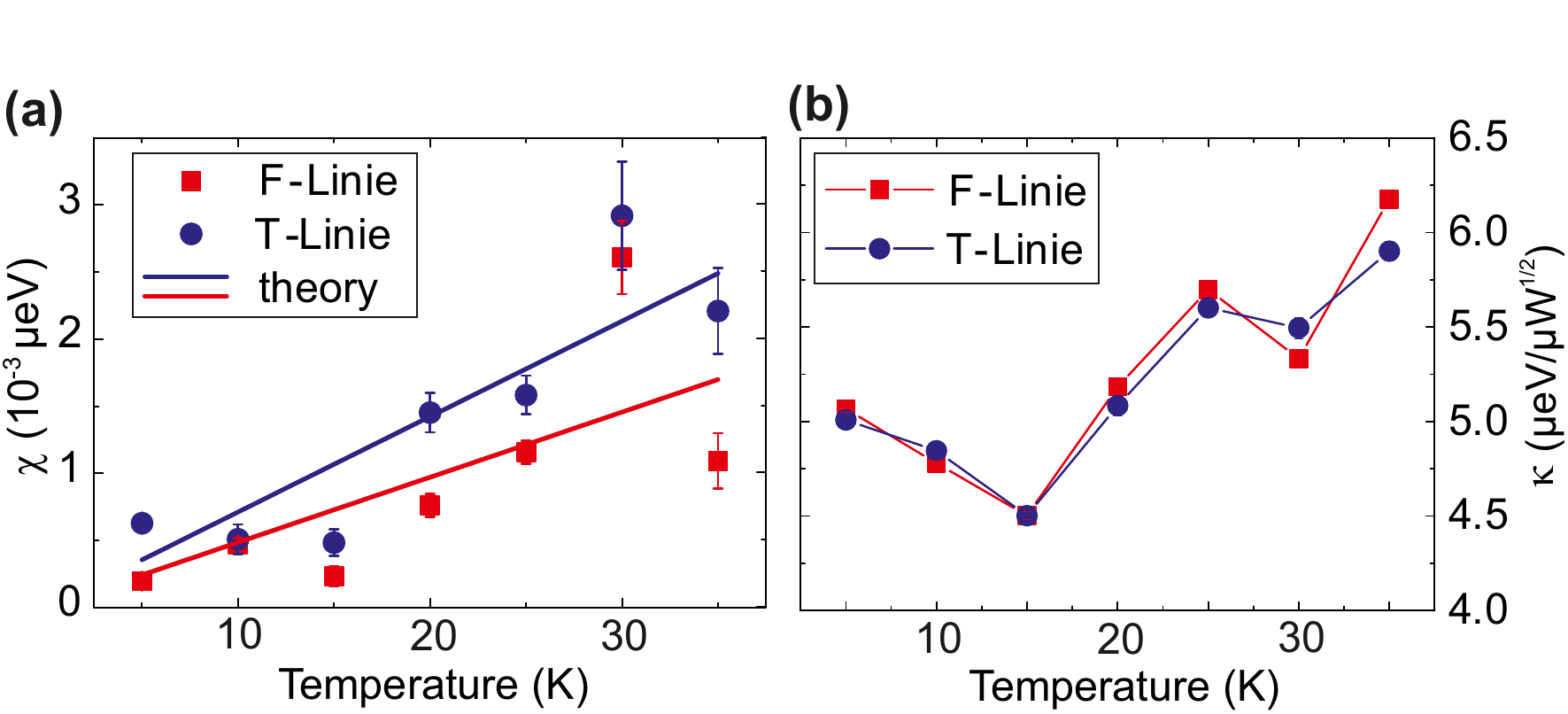}
\caption{(a) Change in the Mollow sideband widths with squared Rabi frequency as a function of temperature. The 
lines show a fitting to the exciton--phonon coupling model. (b)~Gradient of the renormalised Rabi splitting with increasing pump power versus the temperature. We observe a notable increase for temperatures exceeding $15$ K, 
which is attributed to a localization effect of the excitonic wave function for low temperatures.}
\label{Fig4}
\end{center}
\end{figure}

To further characterize our system, we performed measurements of the sidepeak splitting and linewidth for temperatures up to $35~$K. 
In Fig. \ref{Fig4}(a) we plot $\chi$ (see Eq. \ref{gammaPD})  as a function of the sample temperature . $\chi$ is the parameter which captures the power broadening of the mollow sidepeaks. We observe a monotonous increase of $\chi$ with the sample temperature, and by fitting to Eq. \ref{gammaPD} we can extract $\alpha_F=(0.077\pm0.015)~\mathrm{ps}^2$ and $\alpha_T=(0.113\pm0.012)~\mathrm{ps}^2$ which are consistent with other experiments carried out on standard In(Ga)As QDs~\cite{Wei2014,Ramsay2010,Ramsay2010a}. The solid lines in Fig.~\ref{Fig4}(a) show $\gamma_{PD}/\Omega_r^2=\pi\alpha k_B T$ using these extracted values. We attribute the slight scattering of the datapoints to a slight detuning caused by the high driving strength we used in the experiment. As mentioned above, the power dephasing coefficient is a sensitive function to the laser detuning, which we cannot exclude to slightly change during the experiment.

Fig.~{\ref{Fig4}}(b) shows the renormalized dipole moment $\kappa=\mu ~R$ (see equ. \ref{Omegar}) as the temperature is increased. We first obtain a decrease in the effective dipole moment for temperatures up to $15$~K, followed by a clear increase for higher temperatures. From Eq.~({\ref{Omegar}}), we read $\mathrm{d}\Omega_r/\mathrm{d}\sqrt{P}\propto\mu ~R$. Since the phonon renormalisation parameter $R$ decreases with rising temperature, one also expects a decrease of $\mathrm{d}\Omega_r/\mathrm{d}\sqrt{P}$ with temperature. This is what we observe for T<15~K. Above 15~K, the notable increase suggests an increase in the dipole moment $\mu$ with temperature. We assume that the Indium concentration in our QD is non-uniform, which leads to shallow potential minima within QD that trap the exciton at very low temperatures. Once a characteristic activation energy is thermally overcome, the exciton wavefunction spreads over the entire QD, leading to an increase of the dipole moment $\mu$, and thus to a stronger coupling to the resonant laser field. Similar observations were found  via magneto optical studies on self-assembled QDs \cite{Musial2014}.

\begin{figure}
\begin{center}
\includegraphics[width=0.48\textwidth]{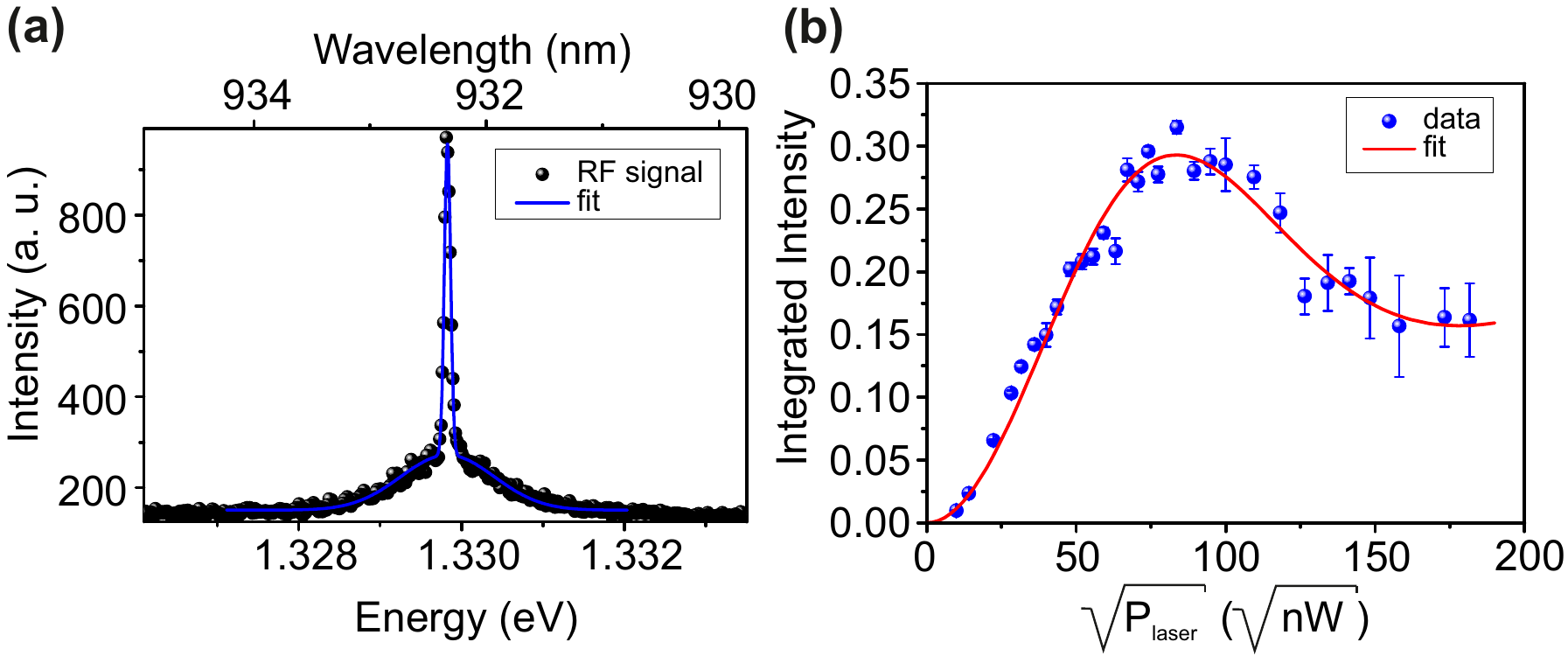}
\caption{(a) Quantum dot spectrum under pulsed resonant excitation and for a sample temperature of $T=5.8$ K. The fit is a double Gaussian function for the dot emission and the remaining 
laser background. (b)~Integrated intensity of the quantum dot emission as a function of the square root of the pump power. The intensity is obtained from fitting the single spectra. The solid line is a fit to theory including coupling to phonons.}
\label{Fig5}
\end{center}
\end{figure}

For potential applications of our SCQDs, it is desirable that the excited state population can be coherently controlled, which we now demonstrate using pulsed resonant excitation. Fig.~\ref{Fig5}(a) shows the emission spectrum of the same SCQD under pulsed resonant excitation (f$_{Rep}=82$~MHz, $\Delta t\approx1.2$~ps). A broad remaining laser background is observed due to the temporally narrow excitation pulses. To extract the quantum dot emission intensity, we fit the QD emission in each single spectrum with a Gaussian function and extract the integrated intensity from the fit results.

The results are shown in Fig.~\ref{Fig5}(b), where a damped sinusoidal behaviour is observed. This is consistent with our exciton--phonon coupling model,  which (according to Eq.~({\ref{gammaPD}})) predicts an increase in dephasing for larger pulse strengths, and has been observed elsewhere for conventional, self assembled QDs~\cite{Ramsay2010a,Ramsay2010,McCutcheon2010}. In fact, we can use Eqs.~({\ref{alphaxdot}}) to ({\ref{alphazdot}}) to fit the experimentally measured intensities as a function of pulse area. The integrated emission intensity is proportional to the final excited state population, which we can simulate using the optical Bloch equations. To do so, we make the replacement $\Omega_r\to R(\Theta/2\tau\sqrt{\pi}) \exp[-(t/2\tau)^2]$, with $\tau=\Delta t/4\sqrt{\ln 2}$, $\Theta$ the pulse area, and we set $\Gamma_1=0$ since we are interested in timescales of the order $\Delta t \ll T_1$. We then numerically solve Eqs.~({\ref{alphaxdot}}) to ({\ref{alphazdot}}) using the full phonon-induced pure-dephasing rate in Eq.~({\ref{gammaPDFull}}) for each excitation power, allowing for a scaling factor between $\sqrt{P_{\mathrm{laser}}}$ and $\Theta$, fixing $\gamma_0=30~\mu\mathrm{eV}$ from our previous fits, and allowing the exciton--phonon coupling parameter $\alpha$ and the cut-off frequency $\omega_c$ to vary.

The result of this fitting procedure is shown by the solid line in Fig.~\ref{Fig5}(b), and the exciton--phonon coupling parameters we extract are $\alpha=(0.054\pm0.016)~\mathrm{ps}^2$ and $\omega_c=(4.34\pm1.21)~\mathrm{ps}^{-1}$. We see that our model is able to capture the experimental data well, though the exciton--phonon coupling strength $\alpha$ we extract differs slightly from those found for continuous wave driving. There are a number of possible reasons for this discrepancy. Firstly, we note that similarly good fits can be found in the pulsed excitation case by fixing $\alpha$ to a value closer to that found for continuous driving, and allowing instead the background dephasing rate $\gamma_0$ to vary. As such, in the case of pulsed excitation it is not straightforward to distinguish phonon-induced dephasing from other sources. We also note that while the fundamental exciton--phonon coupling parameters $\alpha$ and $\omega_c$ should not depend on the excitation scheme, the background rate $\gamma_0$ may well differ in the two cases, particularly for temporally short pulses which have spectral components which are not resonant with the exciton transition energy~\cite{Wei2014a}. Finally, we note that in replacing the Rabi frequency $\Omega_r$ in Eq.~({\ref{gammaPDFull}}) with a time dependent quantity requires making a Markovian approximation, which assumes the phonon environment relaxes on a timescale shorter than the excitonic dynamics~\cite{Ramsay2010,McCutcheon2010}. For the short temporal pulses used here this approximation may not be strictly valid, and we note that it will be interesting to explore potential non-Markovian effects in these systems in future works~\cite{mccutcheon2015}.

\section{Summary}
\label{sec:Summary}

We have shown for the first time the coherent coupling of a resonant laser field to the excitonic state of a SCQD. We observed the characteristic Mollow triplet in the spectral domain and an increase in the Rabi splitting with increasing temperature, which we attribute to a localization of the SCQD wave function for low temperatures due to a gradient in the Indium concentration inside the SCQD. Furthermore, we analyzed in detail the power and temperature dependent dephasing channels in our system. Finally, we demonstrate the coupling of the QD exciton to a pulsed resonant laser field by measuring the dependency of the SCQD emission on the pulse area of the excitation laser. The observed Rabi oscillations  feature a distinct damping, which points towards the presence of exciton--phonon coupling in our system. The possibility to generate resonance fluorescence photons from positioned quantum dots opens new experimental opportunities for realizing scalable solid state quantum emitters. It is a crucial step towards the deterministic generation of single photons with unity indistinguishability \cite{he2013demand}, and resonant techniques are key for obtaining optical coherent control over single spins in QDs \cite{Press2008}. The implementation of such schemes, based on a fully scalable quantum emitter architecture (such as provided by site controlled quantum dots), remains one of the big challenges in solid state quantum emitter research. As a next step towards this architecture, an implementation of site-controlled quantum dots within microcavities is desirable. Due to the control over the nucleation spot, one can expect an almost perfect coupling of the SCQD to the fundamental optical mode of such a device which gives rise to increased extraction efficiencies and high degrees of indistinguishability, which is key for realizing an efficient source of indistinguishable photons.

\section*{Funding Information}
State of Bavaria and the German Ministry of Education and Research (BMBF) within the projects Q.com-H and the Chist-era project SSQN. German Research Foundation via the SFB 787 "`Semiconductor Nanophotonics: Materials, Models, Devices"'.
D.P.S.M. acknowledges support from project SIQUTE (contract EXL02) of the European Metrology Research Programme (EMRP). 
The EMRP is jointly funded by the EMRP participating countries within EURAMET and the European Union.
Y. H. acknowledges support from the Sino-German (CSC-DAAD) Postdoc Scholarship Program.  J. M. acknowledges support from Villum Fonden via the NATEC Centre.

\section*{Acknowledgments}

The authors thank M. Emmerling for sample preparation.


\end{document}